\documentclass[conference]{IEEEtran}
\IEEEoverridecommandlockouts
\usepackage{cite}
\usepackage{amsmath,amssymb,amsfonts}
\usepackage{algorithmic}
\usepackage{multirow}
\usepackage{graphicx}
\usepackage{textcomp}
\usepackage{xcolor}
\usepackage{lipsum,booktabs}
\usepackage{soul}
\usepackage{url}

\usepackage{color, colortbl}
\definecolor{Gray}{gray}{0.9}
\usepackage[first=0,last=9]{lcg}

\def\BibTeX{{\rm B\kern-.05em{\sc i\kern-.025em b}\kern-.08em
    T\kern-.1667em\lower.7ex\hbox{E}\kern-.125emX}}
\begin{document}

\title{Securing an Application Layer Gateway: \\
An Industrial Case Study}

\author{
\IEEEauthorblockN{Carmine Cesarano, Roberto Natella}
\IEEEauthorblockA{\textit{Universit\`a degli Studi di Napoli Federico II, Italy} \\  
\{carmine.cesarano2, roberto.natella\}@unina.it}
}

\maketitle

\begin{abstract}
Application Layer Gateways (ALGs) play a crucial role in securing critical systems, including railways, industrial automation, and defense applications, by segmenting networks at different levels of criticality. However, they require rigorous security testing to prevent software vulnerabilities, not only at the network level but also at the application layer (e.g., deep traffic inspection components). This paper presents a vulnerability-driven methodology for the comprehensive security testing of ALGs. We present the methodology in the context of an industrial case study in the railways domain, and a simulation-based testing environment to support the methodology.
\end{abstract}


\section{Introduction}
The interconnection of diverse applications and networks is essential for building complex systems. Application Layer Gateways (ALGs) are responsible for handling the communication between different networks, with different levels of criticality, and for performing security controls. 
In particular, ALGs are crucial in critical domains such as railways, avionics, and defense. In these domains, network segments for industrial control (e.g., TDRP traffic in the railways domains) should be protected from attacks arising from other network segments (e.g., multimedia traffic over HTTP, FTP, or XMPP); at the same time, communication between network segments should be allowed for specific cases (e.g., for diagnostic purposes). The ALG performs security controls, by analyzing the traffic at the application layer to enforce security policies (e.g., based on the contents and on the authors of the data). 

Given the role of ALGs in such critical domains, there is a need for methodologies for rigorous security testing, in order to prevent software vulnerabilities that could allow attackers to escape security controls. Security standards, such as Common Criteria \cite{commoncriteria}, define security functional requirements (SFR) catalogs for several software categories, including ALGs \cite{security_target}. The same standard also defines the security assurance requirements (SARs) for specific types of software. Although SARs can be useful to assess software security, they are broadly expressed and lack completeness for ALG systems \cite{security_target}. Furthermore, in the literature, no comprehensive methodologies have been proposed for the assessment of ALG systems. Existing toolkits and techniques focus on network layer attacks (e.g., IP and port scanning \cite{anderson2003introduction}, IP and MAC spoofing, or frameworks like BurpSuite \cite{bennetts2013owasp} and ZAP \cite{mahajan2014burp}), leaving out the application layer.

In this work, we present a novel methodology for assessing the security of ALG systems. Our proposal offers a comprehensive approach that addresses not only network-layer vulnerabilities but also delves into the application layer, where ALGs mainly operate. For example, our tests include attack scenarios involving techniques like fuzzing and stress testing specifically tailored to trigger vulnerabilities at the application layer. The contributions of this paper include:

\begin{itemize}
    \item We conduct an in-depth analysis of potential ALG system weaknesses and propose an assessment methodology for their systematic identification.
    \item We develop a simulation-based testing environment for performing ALG systems evaluations, improving the test reproducibility and system observability.
    \item We present a practical case study in the railways domain, demonstrating the application of our methodology to assess the security of an ALG system built on open-source components.
\end{itemize}

The proposed methodology offers systematic guidance for assessing ALG products, covering network and application layers in a holistic testing framework. The methodology is adaptable to different ALG systems, while the practical case study is meant to demonstrate its applicability in real systems. 

In the following of this paper, Section \ref{sec:related} provides an overview of related work concerning ALGs, Section \ref{sec:methodology} presents the proposed methodology, Section \ref{sec:casestudy} introduces the case study, Section \ref{sec:experimental} outlines the experimental analysis, and Section \ref{sec:conclusion} summarizes findings and outlines future research directions.

\section{Related work}
\label{sec:related}
\setlength{\tabcolsep}{0.7\tabcolsep}

\begin{table*}
  \caption{\label{tab:CWE}CWEs and Attack Scenarios in ALGs and Firewall software}
  \setlength{\tabcolsep}{0.7\tabcolsep} 
  \centering
  \begin{tabular}{ p{0.08\linewidth} p{0.43\linewidth} p{0.29\linewidth} p{0.15\linewidth} } 
    \toprule
    \textbf{CWE-ID} & \textbf{Description} & \textbf{Attack Scenario} & \textbf{Occurred in}\\
    \midrule

    CWE-20 & Improper Input Validation & Crafted DNS, RTSP packets & Cisco ASA, Cisco FTD \\ 

    CWE-22 & Improper Limitation of a Pathname to a Restricted Directory & Crafted HTTP requests & HAProxy, Pfsense \\

    CWE-74 & Improper Neutralization of characters ('Injection') & Encapsulation attack (via HTTP req) & HAProxy\\

    CWE-78 & Improper Neutralization of characters ('OS Command Injection') & OS command injection (via HTTP req) & Haproxy, Pfsense \\

    CWE-79 & Improper Neutralization of characters ('Cross-site Scripting') & Crafted HTTP requests & Pfsense\\ 

    CWE-91 & XML Injection (aka Blind XPath Injection) & Manipulation of configuration XML files & Pfsense \\

    CWE-120 & Buffer Copy without Checking Size of Input ('Buffer Overflow') & Specific FTP transfer & Cisco IOS \\

    CWE-190 & Integer Overflow or Wraparound & HTTP Request Smuggling & HAProxy \\

    CWE-200 & Exposure of Sensitive Information to an Unauthorized Actor & Log Data Extraction & HAProxy \\ 

    CWE-281 & Improper Preservation of Permissions & Crafted FTP commands & Pfsense \\
    
    CWE-290 & Authentication Bypass by Spoofing & IP and MAC spoofing & PfSense \\
    
    CWE-307 & Improper Restriction of Excessive Authentication Attempts & Brute Force to Authentication & Pfsense \\

    CWE-358 & Improperly Implemented Security Check for Standard & NAT Slipstreaming & Cisco ASA, Cisco FTD \\

    CWE-399 & Resource Management Errors & Crafted SIP, H.323 packets & Cisco IOS \\

    CWE-400 & Uncontrolled Resource Consumption & Resource Exhaustion Attacks & F5 BIP-IP AFM\\

    CWE-401 & Missing Release of Memory after Effective Lifetime & Crafted SIP packets & Junos OS \\

    CWE-434 & Unrestricted Upload of Dangerous File & Specific FTP transfer & Cisco IOS \\
    
    CWE-444 & Inconsistent Interpretation of HTTP Requests & HTTP Request/Response Smuggling & HAProxy \\

    CWE-459 & Incomplete Cleanup & Crafted HTTP requests & HAProxy \\

    CWE-665 & Improper Initialization & Crafted SIP packets & Cisco IOS\\

    CWE-693 & Protection Mechanism Failure & Crafted H.323 packet & Cisco IOS\\

    CWE-754 & Improper Check for Unusual or Exceptional Conditions & Crafted DNS packets & Cisco IOS \\

    CWE-755 & Improper Handling of Exceptional Conditions & Crafted HTTP request & HAProxy \\

    CWE-787 & Out-of-bounds Write & Crafted HTTP requests & HAProxy \\

    CWE-824 & Access of Uninitialized Pointer & Crafted SIP packets & Junos OS \\

    CWE-835 & Loop with Unreachable Exit Condition ('Infinite Loop') & Crafted HTTP responses & HAProxy \\

    CWE-908 & Use of Uninitialized Resource &  Resource Exhaustion Attacks & F5 BIP-IP AFM \\

    \bottomrule
  \end{tabular}
\end{table*}

Many commercial and open-source systems have been proposed for ALGs. Most notably, GoThings \cite{de2015gothings} proposes an extensible architecture that facilitates interconnectivity between different messaging protocols such as REST, CoAP, MQTT, and XMPP. However, even if these systems address connectivity challenges in the IoT landscape, they are not primarily oriented toward the enforcement of security policies. 

Other products, instead, are specifically designed to ensure security policies in communication. For example, we can mention Multiple Independent Levels of Security (MILS) kernels \cite{alves2006mils}, a high assurance architecture for handling separate execution environments with different classification levels and controlled information flow among them. MILS security guards \cite{mils_guard} are specialized components in secure MILS kernel communication, and can be considered as an ALG.

Sec-ALG \cite{riaz2020sec} prioritizes secure end-to-end communication between private networks. It relies on Deep Packet Inspection for protocol detection and packet filtering. Despite its emphasis on security policies, Sec-ALG does not introduce methodologies for evaluating the effectiveness of its security measures. This underscores the need for a comprehensive testing approach, especially in critical environments where security is a key aspect. 

ALGs aim to impose stringent security policies but can be susceptible to sophisticated attack techniques like BALG \cite{roschke2011balg}, which uses polymorphic and encrypted shellcodes to bypass security measures. However, this attack is highly specific, and evaluating ALGs requires a testing methodology with diverse attack vectors across various layers (e.g., network, application, and policy).

Our paper introduces a comprehensive methodology for the security testing of ALGs deployed in critical domains. This methodology has been designed to address the limitations inherent in existing approaches, offering a systematic evaluation of ALG security, including policy enforcement, network layer, and application layer vulnerabilities.

\section{Methodology}
\label{sec:methodology}
The proposed methodology for assessing ALG security includes four key steps. First, we conduct a thorough analysis of Common Weakness Enumeration (CWEs) relevant to ALG systems. We then define attack scenarios using a CWE-driven approach. Subsequently, we configure a simulated testing environment. Finally, we conclude the security evaluation by executing the tests and analyzing the results.

\subsection{CWE-driven risk analysis}
\label{sec:CWE-analysis}
The proposed methodology is driven by an initial risk analysis phase, which includes an examination of Common Vulnerabilities and Exposures (CVEs) affecting software products within the same category of ALG. Subsequently, we retrieved corresponding CWEs for each identified CVE. 

Table \ref{tab:CWE} presents the results of this phase, using commercial and open-source software as a sample, including ALGs, proxies, and firewalls. For example, \textit{CVE-2021-40346} (\textit{CWE-190}) reveals an integer overflow in HAProxy, enabling attackers to bypass configured Access Control Lists (ACLs). Similarly, \textit{CVE-2019-12655} (\textit{CWE-20}) discloses a buffer overflow, allowing a remote attacker to trigger a Denial of Service (DoS). Due to space constraints, the detailed mapping between CVEs and CWEs are available at \footnote{\texttt{https://dessert.unina.it:8088/carmain/edcc2024}}; only the aggregated CWE results are presented in the Table. The CWEs listed in the table are representative instances of common weaknesses that may affect ALGs. However, they could certainly be augmented with additional analysis from ongoing research on other products or emerging vulnerabilities. This preliminary phase, starting with CVEs and progressing through CWEs, enables the building of a focused knowledge base on potential risks associated with ALGs. Consequently, it guides the phase of defining attack scenarios.

\subsection{Attack scenarios definition}
\label{sec:attack_scenarios}
In the next step of our methodology, we define the security tests required to validate the security posture of ALG. By adopting a CWE-driven approach, we can conduct security testing more focused on real-world weaknesses to which the system may be susceptible. Using the insights gained from the CVE analysis, we can extract details on the exploits associated with each vulnerability, thus facilitating the formulation of the corresponding attack scenario. The fourth column in Table \ref{tab:CWE} enumerates these inferred attack scenarios. For example, according to the CVE description, an HTTP request smuggling attack can be used to assess CVE-2019-12655 mentioned in Section \ref{sec:CWE-analysis}. In the case of CVE-2019-12655, a crafted FTP traffic attack can be employed. By executing all defined attack scenarios, we effectively assess all vulnerabilities we identified during the risk analysis phase.

For each defined attack scenario, a set of testing techniques can be selected. For example, we can select generation fuzzing to generate the FTP traffic mentioned above or mutation fuzzing to mutate benign traffic sniffed from the network to assess weaknesses at the application layer. We can select packet manipulation and crafting tools to assess weaknesses at the network layer. Additionally, we can select stress testing to trigger DoS vulnerabilities. We further detail the employed testing techniques in Section \ref{sec:experimental}.

\subsection{Testing environment setup}
In industrial domains, the assessment process for ALGs typically involves deploying the software component in a real pre-production environment. This is desirable, as the System Under Test (SUT) may exhibit different behaviors depending on the operational environment. The closer these conditions align with those of the production environment, the more effective security testing will be in identifying vulnerabilities that could be exploited when the system is in use. This process is not always feasible and may pose complexities. Conducting ALG testing within the real system can potentially pose risks to costly and safety-critical components. Furthermore, it is not guaranteed that the physical system encompasses debugging tools for observing the SUT's state. Additionally, this approach does not align with the security-by-design principles, which are crucial for security-critical and safety-critical systems, where there is a need to apply security testing from the earliest phases of the design process. To address these challenges, this work proposes the use of a virtual testing environment. Using network simulation technologies, we simulate the network infrastructure within which the SUT will operate, including topology parameters such as subnets, routing configurations, switches, and network protocols, as well as IP addresses, and port configurations. Beyond all this, network simulation facilitates tasks such as monitoring traffic and resources. Furthermore, we use lightweight virtualization and hardware virtualization to emulate the network hosts. During testing, hosts generate and exchange network packets through the ALG. The testing environment also supports the integration of physical machines into the simulated network infrastructure to accommodate non-virtualizable hardware bundled with ALG. This approach is suitable for assessing the security of third-party ALG products, whether open-source or commercial.

\subsection{Security Testing}
In the last phase of our methodology, we implement and perform the security tests according to the defined attack scenarios.  During test execution, we capture and record critical data, including the state of the SUT, system logs, performance metrics, and network traffic using virtualized introspection capabilities. In the final phase of our methodology, we analyze the behavior traces and collected metrics to assess security effectiveness, identify vulnerabilities, and provide recommendations for improvement.

\section{Case study}
\label{sec:casestudy}
\begin{figure}
  \includegraphics[width=\linewidth]{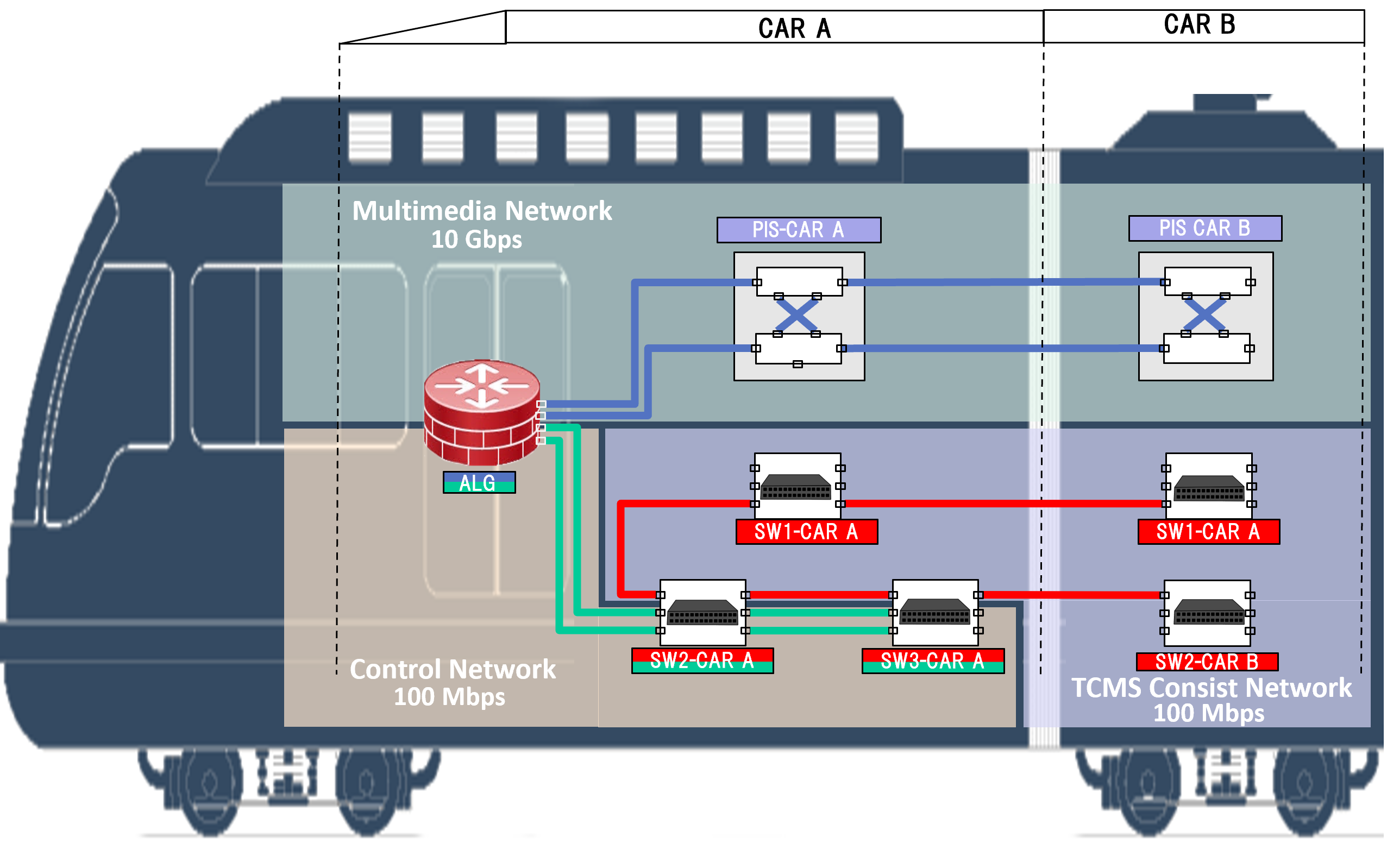}
  \caption{On-board Train Network}
  \label{fig:on_board_network}
\end{figure}

In order to demonstrate the practical application of the proposed methodology we present a case study involving the integration of an ALG into a real-world network infrastructure, specifically an On-Board Train Network. The analyzed case study stems from a collaboration with an Italian railways company.

\subsection{On-board Train Network}
\label{sec:topology}
Figure \ref{fig:on_board_network} shows the network topology under analysis. All on-board services and functions are segregated from each other using VLANs, as described by the EN 50159 standard \cite{EN_50159}, which governs communication, signaling, and processing in safety-related transmission systems. This segmentation is essential to ensure the safety and reliability of the train's operations. The on-board VLANs can be categorized into three main groups: 1) VLANs for the Train Control and Monitoring System (\textit{TCMS}), 2) VLANs for train control services (maintenance, audio, on-board intranet, train integrity, fire protection), and 3) VLANs for passenger multimedia services. In each VLAN, there are one or more switches to which the hosts are connected (CCTV cameras, IP cams, PCN sensors, loudspeakers, etc.).

VLANs must be isolated, but at the same time some communications are allowed. For this reason, the ALG is critical in ensuring the security of the communication between VPNs with different levels of criticality. For example, the train control service VLAN can transfer information to the passenger multimedia service VLAN, and vice versa.

\begin{table}
  \caption{\label{tab:security_policies}Security Policies Categories}
  \setlength{\tabcolsep}{0.7\tabcolsep}
  \centering
  \begin{tabular}{ p{0.80\linewidth} p{0.15\linewidth}}
    \toprule
    \textbf{Security Policy} & \textbf{Type} \\
    \midrule

    \rowcolor[gray]{.9}
    \textbf{Network Isolation}: Block all direct traffic between the two networks unless explicitly authorized & Firewall \\

    \rowcolor{white}
    \textbf{IP Filtering}: Allow traffic only from authorized IP addresses in the low-criticality network. & Firewall \\ 

    \rowcolor[gray]{.9}
    \textbf{Ports and Protocols}: Allow only traffic based on specific protocols (e.g., HTTP, HTTPS, FTP) and block unnecessary protocols. & Firewall \\ 

    \rowcolor{white}
    \textbf{User/Role Access}: Allow access to sensitive resources in the high-criticality network only for authorized users based on their roles. & Application\\ 

    \rowcolor[gray]{.9}
    \textbf{Content Inspection}: Apply Deep Packet Inspection to identify malicious payloads or anomalous behavior. & Application\\ 

    \rowcolor{white}
    \textbf{URL Filtering}: Block access to URLs or domains known to be unauthorized. & Application\\ 

    \rowcolor[gray]{.9}
    \textbf{Protocol-based and service-based routing}: Route traffic based on the specific protocol or type of service to its designated servers in the high-criticality network. & Routing  \\ 

    \rowcolor{white}
    \textbf{Bandwidth Limitations}: Limit the available bandwidth for certain traffic categories to ensure fair use of network resources. & Routing \\

    \bottomrule
  \end{tabular}
\end{table}






\subsection{ALG security policies}
\label{sec:security_policies}
The ALG deployed in the on-board network mitigates security issues, including information disclosure, data integrity, and DoS, by controlling traffic between VLANs of varying levels of criticality. The ALG intercepts and inspects each packet, applying policies to enforce security between the VLANs.

The first responsibility of the ALG is to provide firewall capabilities, operating at the network level (Layer 3). Policies for filtering packets can be based on IP addresses, ports, specific protocols, and service types. ALG can restrict traffic from a specific set of source IPs and allow or block certain protocols (e.g., HTTP and FTP) or prevent access to services (e.g., SMTP and POP3).

The second responsibility of the ALG is to enforce the application level (Layer 7), securing the communication between hosts with different privilege levels. For instance, the ALG can manage requests for privileged resources using HTTP or FTP, maintaining a whitelist of authorized URLs. Policies based on users or roles are enforced, allowing access only for authorized users. Deep packet inspection is employed to identify malicious payloads containing malware.

Additionally, the ALG provides separation between different VLANs, ensuring unawareness of each other's network topology and composition. Therefore, the third responsibility of the ALG is to route traffic to the correct destination host based on the requested protocol or service type. 

For each of the ALG's responsibilities, security policies falling under different categories can be configured. Table \ref{tab:security_policies} provides a list of these categories, specifying the types including firewall, application, or routing.

\section{Experimental Analysis}
\label{sec:experimental}

\begin{figure}
  \includegraphics[width=\linewidth]{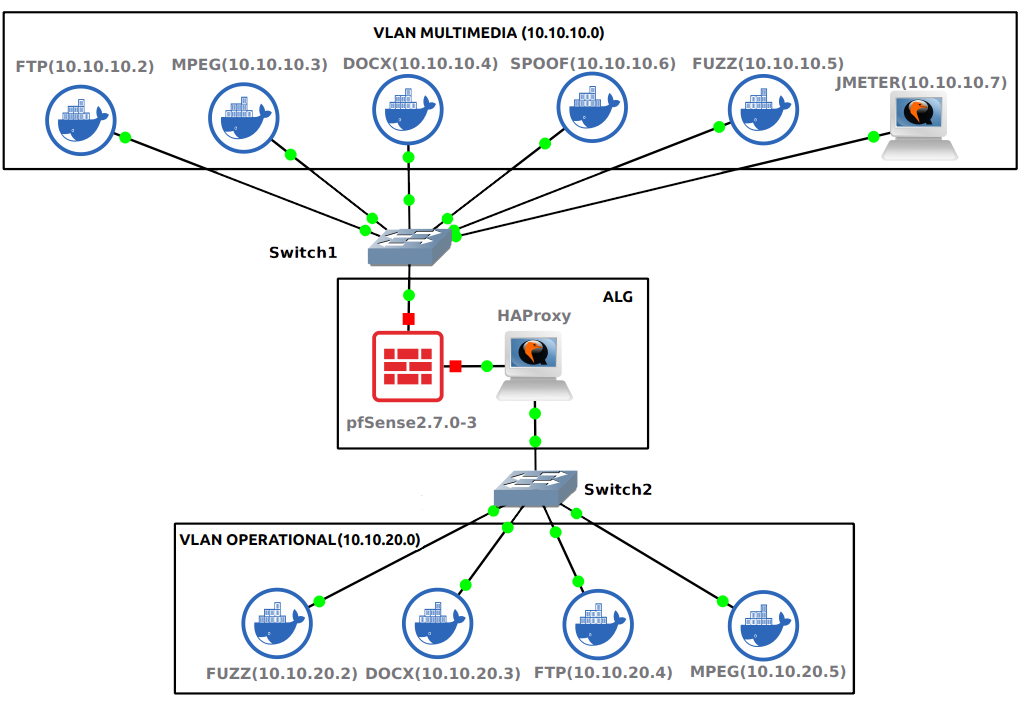}
  \caption{On-board train network: simulation on GNS3}
  \label{fig:network_simulation}
\end{figure}

In this section, we will present the simulated network setup, the deployment and configuration of an open-source ALG, and its security assessment by using the proposed methodology.

\begin{table*}
  \caption{\label{tab:test_performed}ALG Security Policies and Attack Scenarios}
  \setlength{\tabcolsep}{0.7\tabcolsep} 
  \centering
  \begin{tabular}{ p{0.08\linewidth} p{0.38\linewidth}  | p{0.32\linewidth} p{0.17\linewidth} } 
    \toprule
    \rowcolor{white}
    \textbf{CWE} & \textbf{Security Policy} & \textbf{Attack Description} & \textbf{Result} \\
    \midrule

    \rowcolor[gray]{.9}
    CWE-20 & ALG shall manage all communications between VLANs. Hosts cannot communicate with each other. & Verify that all and only the configured connections are allowed using ICMP requests & Enforced \\ 

    \rowcolor{white}
    CWE-290 & ALG shall allow only authorized source IPs from a whitelist. & Crafted packets with spoofed source IP as an existing host in the multimedia VLAN. & Not Enforced \\

    \rowcolor[gray]{.9}
    CWE-290 & ALG shall allow only authorized source MACs from a whitelist. & Crafted packets with spoofed source MAC as an existing host in the multimedia VLAN. & Not Enforced \\

    \rowcolor{white}
    CWE-290 & ALG shall ensure static MAC/IP bindings in the ARP table preventing any manipulation. & Crafted ARP packets (poisoning attack) & Enforced \\ 
    
    \rowcolor[gray]{.9}
    CWE-281 & ALG shall block 'MKD' command to the FTP server 10.10.20.5 & Attempt to execute 'MKD' command from the FTP client 10.10.10.2 & Enforced \\

    \rowcolor{white}
    CWE-434 & ALG shall check malicious patterns in data transferred through the 'CP' command to the FTP server 10.10.20.5 & Crafted file sent by the FTP client 10.10.10.2 & Not Enforced\\
    
    \rowcolor[gray]{.9}
    CWE-281 & ALG shall authorize only MPEG files from client 10.10.10.3 through port 8085 and redirect them to server 10.10.20.5 & Crafted HTTP requests with different file types on different ports. & Enforced \\

    \rowcolor{white}
    CWE-281 & ALG shall authorize only DOC files from client 10.10.10.4 through port 8080 and redirect to server 10.10.20.3. & Crafted HTTP requests with different file types on different ports. & Enforced \\

    \rowcolor[gray]{.9}
    CWE-281 & The ALG shall authorize files included in an HTTP payload only if it is authored by a user in a whitelist. & Crafted HTTP requests containing files with random author & Enforced  \\

    \rowcolor{white}
    CWE-444 & ALG shall normalize and standardize incoming HTTP requests to prevent inconsistencies that attackers might exploit. & Crafted HTTP requests & Not Enforced \\

    \rowcolor[gray]{.9}
    CWE-1333 & ALG shall employ an efficient regex for the check of packets, to minimize resource consumption and mitigate ReDoS. & Crafted HTTP requests & Not Enforced \\ 

    \rowcolor{white}
    CWE-400 & ALG shall ensure no performance degradation under normal conditions & Stress testing & No performance degradation within 100K req/min. \\

    \rowcolor[gray]{.9}
    CWE-400 & ALG shall introduce an acceptable maximum latency. & Stress testing & Latency of 39 ms \\ 
    
    \bottomrule
  \end{tabular}
\end{table*}

\subsection{Network and ALG Setup}
\label{sec:ALG_setup}
We employ state-of-the-art GNS3 \cite{gns3} software, which provides an effective platform for network simulation \cite{chou2016comparison}, to build the testing environment. For the sake of simplicity, experiments are conducted on a simplified version of the topology shown in Section \ref{sec:casestudy}, which only includes the \textit{Control Network} VLAN (subnet 10.10.10.0) and the Multimedia Network VLAN (subnet 10.10.20.0). 

Adhering to the on-board network project specifications, hosts within VLANs communicate using various protocols. Open-source protocols like HTTP and FTP are included in the simulation, while closed-source ones, such as TRDP, are deliberately omitted. To simulate the required communications, an FTP server and two HTTP webservers, responsible for FTP connections, MPEG data streams, and DOC file transfers, are instantiated in the Control VLAN. Corresponding clients reside in the multimedia VLAN. Additional hosts, namely \textit{Spoof}, \textit{Fuzz}, and \textit{Jmeter}, are instantiated for security testing operations. All network hosts are instantiated as Docker containers, except for the \textit{JMeter} host, utilizing hardware virtualization for enhanced performance. Figure \ref{fig:network_simulation} provides an overview of the entire simulated network and its interconnected hosts.

We simulate ALG functionalities and security policies by using a set of open-source software components since the available ALGs are commercial products (such as BIG-IP Advanced WAF \cite{bigip} or Palo Alto Networks NGF \cite{paloalto}). We employ Pfsense \cite{pfsense} firewall for the routing layer and HAProxy \cite{haproxy} load balancer for the packet filtering at the application layer. In addition, since HAProxy does not provide advanced packet inspection capabilities, we extended its functionality using some Python custom plugins. The first plugin is designed to handle HTTP requests containing textual files, such as \textit{docx} documents. All HTTP traffic containing documents is redirected to the plugin, which extracts the document and applies security checks to its fields and content. The second plugin extends ALG functionalities for the FTP protocol, applying security checks to the FTP control and data packets. When properly configured, these components simulate the behavior of a real ALG. We deploy the simulated ALG on top of a Linux-based virtual machine connected to a third VLAN (subnet 10.10.11.0) configured with 1 vCPU and 1 GB RAM. After deploying the ALG within the simulated network, we configured multiple security policies from the categories described in Table \ref{tab:security_policies}. 

The first two columns of Table \ref{tab:test_performed} provide details on all configured security policies and their corresponding mitigated CWEs. For example, one of these forces all traffic exchanged between the control VLAN and the multimedia VLAN to pass through the ALG and deny direct connections between hosts. Additionally, IP/MAC whitelisting to reject packets from unrecognized hosts, and IP/MAC binding to prevent ARP cache poisoning are employed. Furthermore, other policies allow the HTTP web servers to handle only specific types of files (i.e., DOC and mpeg), and force routing based on the HTTP content-type field. For example, only HTTP requests containing DOC files are allowed on ALG's 8080 port and redirected to the \textit{DOC} server; requests containing mpeg files are allowed on port 8085 and redirected to the \textit{MPEG} server. It is also required that some FTP commands be not allowed, such as '\textit{mkd}' command.

Other configured security policies address the application layer. For example, ALG checks the HTTP payload or header, allowing only requests containing files created by whitelisted entities. Other ALG policies prevent inconsistencies in the HTTP request format or enforce scanning for malicious patterns in FTP-transferred data. In addition, efficient regex rules must be employed in these checks to minimize resource consumption and mitigate ReDoS attacks.

Lastly, some policies are related to the availability and responsiveness properties of the ALG. For example, ALG shall ensure no performance degradation under normal conditions and shall guarantee an acceptable response time.

\subsection{Experimental results}
In the experimental analysis, we conducted the final phase of our methodology, focusing on the security testing of the deployed simulated ALG. Each security policy, configured to mitigate specific CWEs, underwent assessment for correct enforcement through a specific attack scenario, as outlined in Section \ref{sec:CWE-analysis}. All the performed attacks are listed in Table \ref{tab:test_performed} along with the corresponding outcomes, indicating whether the security policy was '\textit{Enforced}' or '\textit{Not Enforced}'.

We employed various security testing techniques for the implementation of attack scenarios. We used ICMP Echo requests to test whether the ALG allows or denies communication between hosts in separate VLANS. This test passed, confirming that traffic can only pass through the ALG. Then, we use the Scapy tool \cite{rohith2018scapy} to assess other security policies at the network layer and related CWEs. This tool allows us to craft custom packets and design multiple attack scenarios, which we run from the \textit{SPOOF} client on the multimedia VLAN. For instance, we checked if ALG only allows communication from recognized hosts, determined by their IP or MAC addresses. While ALG does deny communication from unrecognized hosts, we discovered that it is vulnerable to IP or MAC spoofing when we send packets using spoofed IP or MAC addresses of recognized hosts. To prevent such attacks, an ALG should incorporate authentication mechanisms, such as the use of certificates.

Furthermore, we executed an ARP cache poisoning attack on the ALG aiming to alter an existing IP-MAC binding in the ALG's ARP table. This manipulation leads the ALG to redirect traffic meant for a victim host in the VLAN to the malicious host (SPOOF client). We crafted an ARP request for the ALG, using the source IP of the victim host and the MAC address of the malicious host. The test has passed because the ALG did not update its ARP table.

To assess application layer security policies, we perform fuzzing tests from the FUZZ client in the multimedia VLAN. Generative fuzzers, such as the BooFuzz tool \cite{boo_fuzz} can be configured to automatically generate HTTP request packets. These tools employ a generation model to define request templates and vary packet parameters, to identify how the target application responds to different inputs. By altering the content-type and payload in the generated packets, we verified that the ALG correctly allows only specific preconfigured file types (e.g., MPEG and DOC) and authorized file authors. 

In addition, BooFuzz discovered two vulnerabilities (CVE-2023-25950 and CVE-2023-25725) in Haproxy. We used a generation model to craft HTTP requests with duplicated Content-Type or Content-Length fields. Normally, the ALG should reject these requests, but it mistakenly processes them, leading to a vulnerability known as \textit{HTTP Request Smuggling}. ALG treats the last duplicated field as valid, causing the generation of multiple responses to the client, and leading to a vulnerability known as \textit{HTTP Response Splitting}.

Mutation fuzzers, such as Mutiny \cite{mutiny}, are also utilized during attack scenarios. We intentionally introduced an inefficient regular expression for malicious pattern recognition in the FTP parser plugin. Mutiny allows us to use FTP sniffed traffic and mutate it to stress the configured regex. This causes the ALG to consume excessive computational resources when processing requests, making the ALG vulnerable to a ReDoS attack.

We conducted additional tests to assess the performance impact of the ALG when processing packets exchanged between two VLANs. Using JMeter, we generated an increasing number of requests from the JMeter client in the multimedia network to the ALG. For the hardware configuration used, we did not observe performance degradation up to 100,000 requests per minute. Beyond this value, we observed linear degradation in terms of response time. Another test involved measuring the latency introduced by the ALG. We again used JMeter to generate HTTP requests and measure response times. By configuring the ALG to forward a response following each request, we could approximate the latency introduced by the ALG, which amounted to approximately 39 ms for the given hardware configuration.

\section{Conclusion and Future Work}
\label{sec:conclusion}
The proposed methodology offers the advantage of being systematic and comprehensive. It provides a clear understanding of specific vulnerabilities affecting ALGs, which can be effectively addressed with tests defined through a CWE-driven approach. This methodology serves as a driving framework for testing and potentially certifying these products, streamlining assessments in critical industrial infrastructures.

In addition, it covers both the network and the application layers within a unified testing environment. Unlike solutions tailored to specific target systems, our approach can be readily adapted to a wide range of systems, making it generalizable and reusable. The provided practical case study further demonstrates the applicability of our methodology in real-world scenarios.

In future work, we plan to use more advanced fuzzing frameworks like AFL, or machine learning-based testing tools to assess other security policies not addressed in this work. We will also extend the experimental analysis, using the proposed methodology for the assessment of other products.

\vspace{5 pt}
\section*{Acknowledgment}
This work was partially supported by the \textit{FLEGREA project} (CUP E53D23007950001) funded by MUR and the \textit{GENIO project} (CUP B69J23005770005) funded by MIMIT.

\vspace{10 pt}
\IEEEtriggeratref{27}
\bibliographystyle{IEEEtran}
\bibliography{bibliography}

\end{document}